\documentclass[12pt,onecolumn]{IEEEtran}
\usepackage{graphicx}
\usepackage{amssymb}
\usepackage{amsmath}
\usepackage{amssymb}
\usepackage{array}
\usepackage{xy}
\usepackage{subfigure}

\begin{document}

\title{Experience versus Talent Shapes the Structure of the  Web}

\author{Joseph S. Kong$^*$\footnote{$^*$Department of Electrical Engineering,
University of California, Los Angeles, CA 90095. Email: jskong@ee.ucla.edu}, Nima
Sarshar$^{\dag}$\footnote{$^{\dag}$Faculty of Engineering, University of Regina, Regina
SK, Canada S4S 0A4 e-mail: nima.sarshar@uregina.ca}, and Vwani P. Roychowdhury$^{\ddag}$
$^{\S}$\footnote{$^{\ddag}$Department of Electrical Engineering, University of California,
Los Angeles, CA 90095. \& Netseer Inc., 11943 Montana Ave. Suite 200, Los Angeles, CA
90049. Email: vwani@ee.ucla.edu}~\footnote{$^{\S}$To whom correspondence should be
addressed.} }

\maketitle
\begin{abstract}
We use sequential large-scale crawl data to empirically investigate and validate the
dynamics that underlie the evolution of the structure of the web.  We find that the
overall structure of the web is defined by an intricate interplay between experience or
entitlement of the pages (as measured by the number of inbound hyperlinks a page already
has), inherent talent or fitness of the pages (as measured by the likelihood that someone
visiting the page would give a hyperlink to it), and the continual high rates of birth and
death of pages on the web. We find that the web is conservative in judging talent, and the
overall fitness distribution is exponential, showing low variability. The small variance
in talent, however, is enough to lead to experience distributions with high variance: the
preferential attachment mechanism amplifies these small biases and leads to  heavy-tailed
power-law (PL) inbound degree distributions over all pages, as well as, over pages that
are of the same age. The exponential distribution of fitness is also key in countering the
potentially destabilizing effect of removal of pages: it stabilizes the exponent of the PL
to a low value, and preserves the heavy tail and the resulting hierarchy, even in the face
of very high rates of uniform deletion of web pages. The  balancing act between experience
and talent on the web allows newly introduced pages with novel and interesting content to
grow fast and catch up or even surpass older pages who have already built their web
presence. In this regard, it is much like what we observe in high-mobility and
meritocratic societies: people with entitlement continue to have access to the best
resources, but there is just enough screening for fitness that allows for talented winners
to emerge and join the ranks of the leaders. Finally, the estimates of the fitness of
webpages and their distribution have potential practical applications in ranking search
engine query results, which can allow users easier access to promising web pages that have
not yet become popular.
\end{abstract}

\section{Introduction}

We, at both the individual and societal levels,  have to constantly make decisions on how
we should distribute our limited resources and time. We need to make choices as to who to
hire, elect, buy from, get information from, award grants to, or make friends with. In
this competitive landscape, each candidate  touts a resume highlighting \emph{experience}
-- a more easily quantifiable metric that summarizes past achievements, e.g., the total
number of clients a service provider has served, or the years a prospective employee has
spent at similar jobs,-- and \emph{talent} or inherent \emph{fitness} -- a more subjective
metric that indicates how well the candidates might perform in the future, e.g., especial
pedigree or degree from a prestigious college, or knowledge of a brand new technology,  or
an articulation of an ideal that captures the imagination. How we strike a balance between
entitlement/experience and fitness/potential is a key determining factor in how wealth and
power get distributed in a society, and how nimble it is in adapting to changes. Too much
emphasis on experience alone could lead to an ossified social structure that lacks
innovation and can collapse dramatically when confronted with change; the world history is
littered with numerous instances of failed societies who had chosen such a path. The
opposite extreme of letting only promising upstarts rule, can equally easily lead to a
state of anarchy with no dominant institutions to hold the society together; the frequent
failures of well-intentioned revolutions that supplant existing institutions en masse and
make fresh starts, provide eloquent testimonies to the perils of such a path. A
society-wide quantitative study of how the experience vs. talent question is resolved,
however, has been difficult to perform because of the obvious lack of concrete data.

The World Wide Web (WWW) provides a unique opportunity in this regard. It has emerged as a
symbiotic socioeconomic entity, enabling new forms of commerce and social intercourse,
while being constantly updated and modified by the activities that it itself enables.
Given the web's organic nature, its evolution, structure, and information dynamics should
reflect many of the same dynamics that underlie its real-world counterparts, i.e., our
social and economic institutions. Thus, we ask how does this thriving cyber-society deal
with the experience vs. talent issue, and how this interplay influences its own structure.
The unprecedented scale and transparency of the activities on the web can provide data
that hitherto has been unavailable. The web is typically modeled as an evolving network
whose nodes are web pages and whose edges are URL links or hyperlinks.  A web page's
in-degree (i.e., the number of other pages that provide links to it) is a good
approximation to its ability to compete, since heavily linked web documents are entitled
to numerous benefits, such as being easier to find via random browsing, being possibly
ranked higher in search engine results, attracting higher traffic and, thus, higher
revenue through online advertisements. Thus {\em the degree of a node} can be considered
{\em as a proxy of its experience}, and it is a reflection of its entitlement, status and
accomplishments to date.

In fact, motivated by a power-law (PL) distribution of the degree of nodes in the web
graph (i.e., $P(k) \propto k^{-\gamma}$, where $k$ denotes node degree and $\gamma$ is the
power law (PL) exponent), the principle of preferential attachment (PA), known to
sociologists and economists for decades (e.g., as the ``cumulative advantage'' or the
``rich gets richer'' principle\cite{simon,price}), was proposed as a dominant dynamic in the web
\cite{barabasi,huberman,huberman_tc}. Note that \cite{huberman} modeled web growth in
terms of growth in the sizes of web sites/domains, which is identical to the model used by
Willis and Yule \cite{yule} in 1922 to explain the PL in the sizes of the genus. However, as shown in
\cite{simkin}, the Yule's model and the Simon's model \cite{simon} are equivalent to each
other, and both rely on the cumulative advantage principle. Hence, we refer to both the
models introduced in \cite{barabasi} and in \cite{huberman,huberman_tc} as the PA model.
Alternate local dynamical models of the web, e.g., via copying of links\cite{kleinberg}
(again, inspired by analogous social dynamics, such as referral services), account for
additional characteristics of the web graph, such as high clustering coefficients and
bipartite clique communities, while still retaining the global PA mechanism.

The PA or equivalent models, however, imply that the scale is heavily tilted towards
experience: the more experienced or older a page is, the more resources it will get and
the more dominant it will become. For example, PA predicts that almost all nodes with high
in-degree are old nodes (disallowing newcomers to catch up), and  that the degree
distribution of pages introduced at the same time will be an exponential one, with very
low variance. This extreme bias of the model was quickly realized and \cite{huberman_tc}
presented empirical data showing that the degree distribution of nodes of the same age has
a very high variance; they also introduced a fitness or talent parameter allowing
different domains to grow at different rates to theoretically account for the high
variance \cite{huberman}. This also prompted a number of researchers to propose
\cite{bianconi,bianconi_be} and explore \cite{borgs,motwani} the ``preferential attachment
with fitness'' dynamical model in which a node $i$ acquires a new link with probability
proportional to $k_i\times \eta_i$, the product of its current number of links $k_i$ and
its intrinsic \emph{fitness or talent} $\eta_i$. In such a linear fitness model, the
degree distribution and the structure of the resulting network \emph{depends on the
distribution of the talent parameter}, $\eta_i$, and thus, {without any knowledge of the
exact distribution, one cannot quite say how exactly the talent vs. experience issue gets
played out in the system}. For example, a uniform distribution of talent has a very
different implication than say an exponential distribution. Moreover, a significant
potential dynamic that has not been studied in the context of the web is the death or
deletion dynamic, which is dominant in most societal settings, where institutions and
individuals cease to operate. The deletion dynamic, however, has been studied in the
context of other networks \cite{sarshar,chung,cooper,moore}, and a surprising finding is
that the heavy-tailed degree distribution disappears in the straight PA model under
significant deletion.

This prompted us to ask \emph{data-driven questions}, such as: How dominant is the churn
or deletion dynamic in the web? Can a PA model with fitness preserve the heavy tail even
in the presence of high deletion rates? Can one empirically verify that the proposed
models are \emph{truly at work} in the web? Can one empirically estimate the relative
fitness of a significant number of pages on the web and quantify the distribution of
talent on the web?  Most interesting of all, how often can talents overtake the more experienced
individuals and emerge as the {\em winner}?  
Such issues, while have been partially theorized about, have not been empirically studied and validated.
\\ \\
\noindent{\bf Brief Summary of Findings. }
Using web crawls that span the period of one year (i.e., 13 separate crawls, at one month
interval), we tracked both the death and the growth processes of the web pages. In
particular, we tracked 17.5 thousand web hosts, via monthly crawls,  with each crawl
containing in excess of 22 million pages (see {\it Materials and Methods}). First, we
discovered that there is a high turnover rate, and for every page created on the web, our
conservative numerical estimates show that at least around $0.77$ pages are deleted (see
{\it Results} and Supporting Information). This is a significant
enough deletion rate that it prompted us to analyze a theoretical model that integrates
the deletion process with the fitness-based preferential attachment dynamics (see
{\it Materials and Methods}). Previous models of the web had neglected the death dynamic;
recent results, however, show that even a relatively low-grade deletion dynamic could
alter network characteristics considerably. Given the distribution of fitness, our model
can predict the overall degree distribution and the degree distribution of nodes with
similar age.

The empirical crawl data is then used to estimate the parameters of the model. This allows
us to validate for the first time whether detailed time-domain data is consistent with the
predictions of the theoretical model. One of the most important assumptions of the model
is that each page can be assigned a constant fitness (which can vary from page to page)
that determines the rate at which it will accumulate hyperlinks. We perform an estimation
of the fitness factor for each month, and show that for the period of the crawls, 
the data do not reject the hypothesis that each page has a constant fitness
(see Supporting Information). A further verification of the model is obtained by validating
one of its most direct implications. In particular, the dynamical model predicts that the
accumulated in-degree (i.e., counting all hyperlinks, including those made by pages that
get deleted during our study period) of a page grows as a power-law. We find that for a
vast majority of pages that show any growth, the degree-vs-time plots in the $\log$-$\log$
scale have linear fits with correlation coefficients in excess of $0.9$ (see the {\it Results} section).
The slope of the linear fit is an affine function of the
fitness of the page.

The robust estimation of the fitness factors of individual pages allows us to determine
the overall distribution.  We find the fitness on the web to be exponentially distributed
(i.e., see Figure~\ref{fig:web_fit_log}(a)), with a truncation. When inserted into our
analytical model, this \emph{exponential fitness distribution correctly predicts} the
power-law degree distribution empirically observed in the overall web as well as for the
set of nodes with similar age. For pages with similar age, the initial exponential
distribution of fitness gets amplified by the PA mechanism, and as a result, the degree
distribution of pages of the same age is a PL distribution with exponent $2$, i.e., with
high variance. Moreover, the truncated exponential distribution of fitness is one of the
few distributions that would generate a constant PL exponent in the overall degree
distribution, even as the turnover rate approaches unity (i.e., as many pages are deleted
as created on the average). The empirical data agrees with this prediction and the PL
degree distribution retains a constant low-magnitude exponent throughout the period of our
study (see the {\it Results} section) even though the deletion rate of pages
remains high. Thus the fitness distribution of the pages helps in preserving the
heavy-tailed scale-free overall degree distribution of the web.

\begin{figure}[htb]
\centering
\includegraphics[width=3.2in]{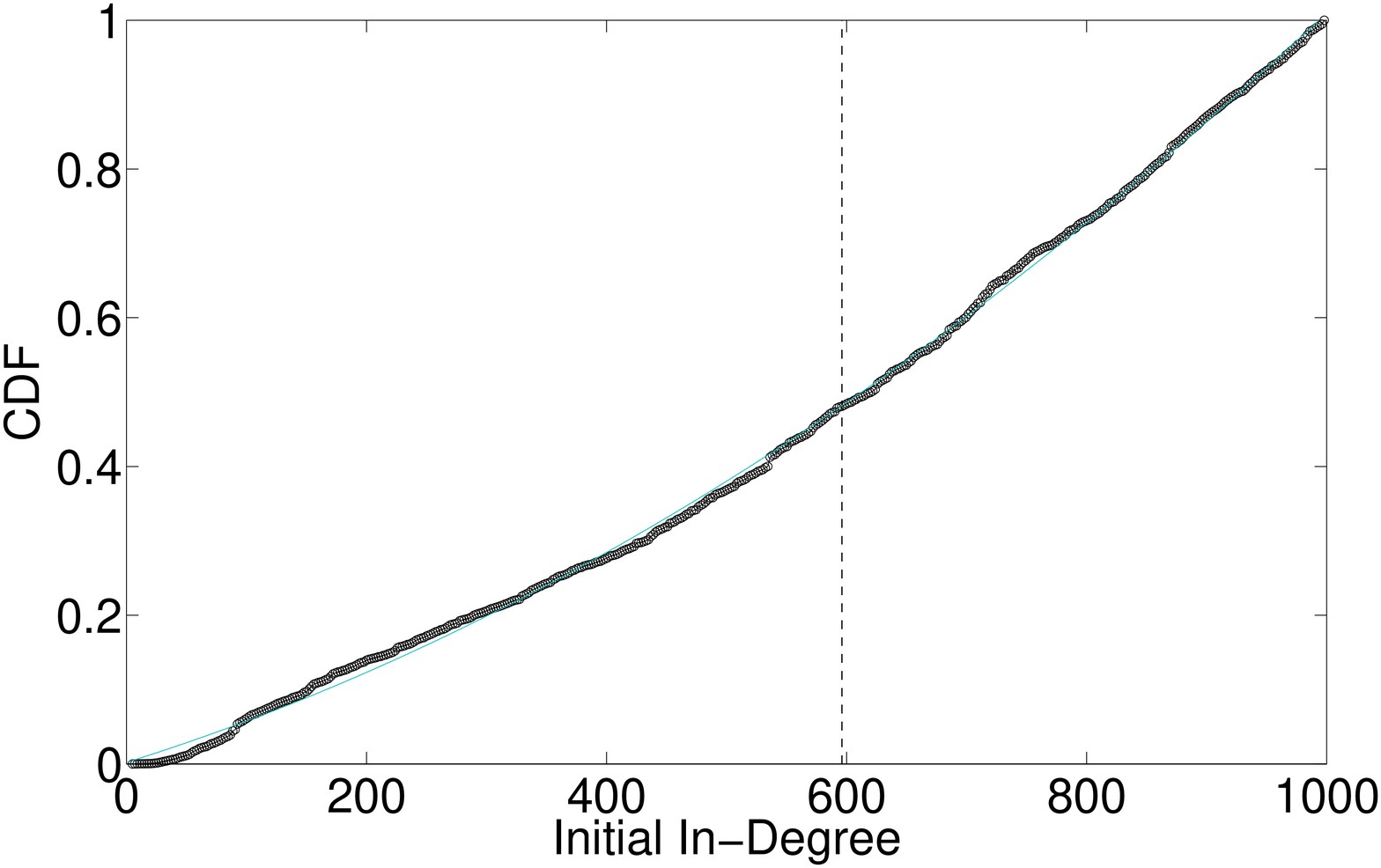}
\includegraphics[width=3.2in]{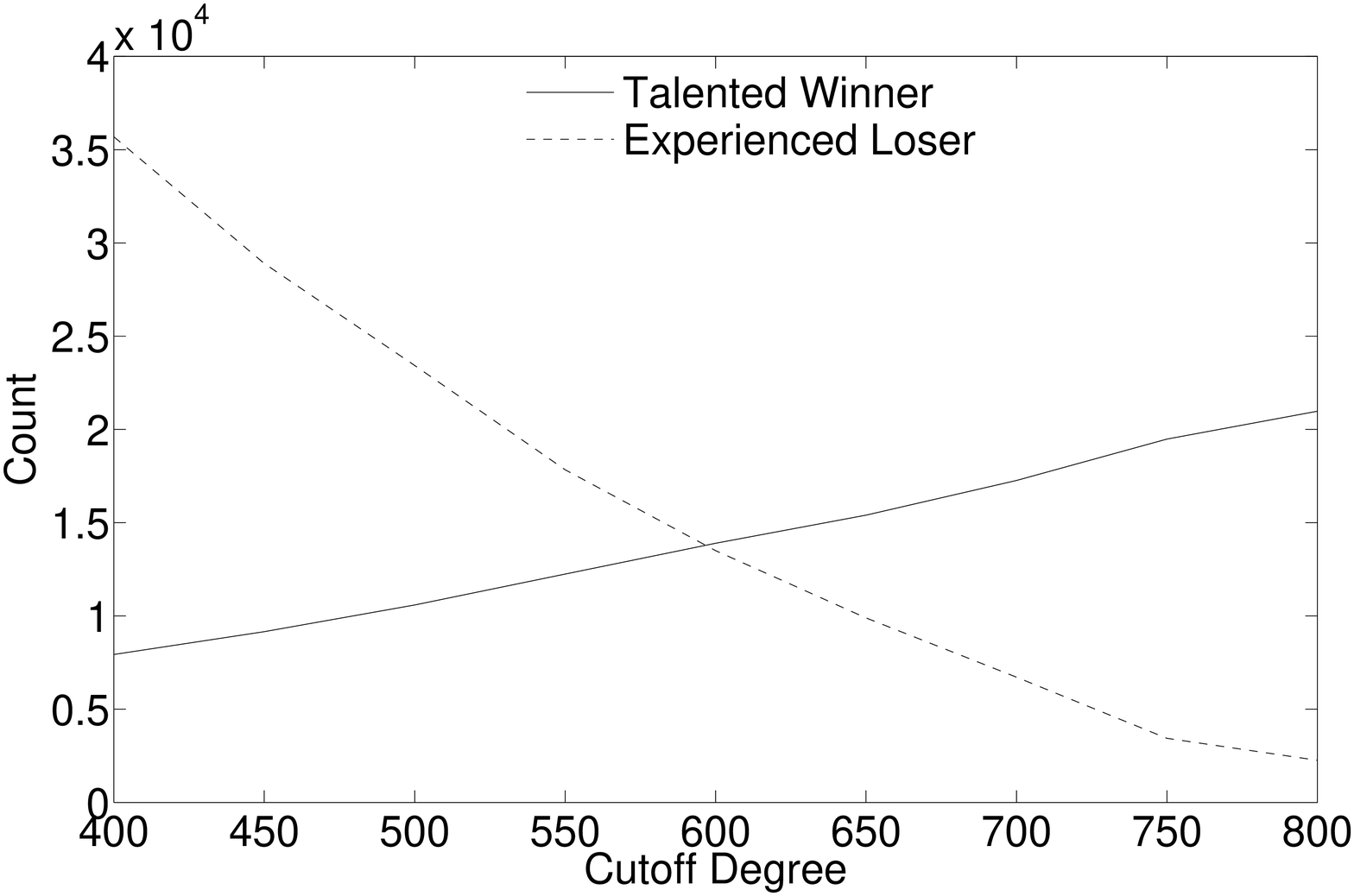}
\caption{\small (a) Cumulative distribution of the initial in-degree (or experience) of
pages with measurable fitness (see the {\it Results} section for the definition of
such pages) \emph{with} the following properties: (i) Initial in-degree in June 2006 was
less than one thousand; (ii) The final accumulated in-degree at the end of the observation
period (i.e., in June 2007) was greater than one thousand. (b) Count vs. cutoff degree:
the downward-slope line denotes the number of \emph{experienced losers} (i.e. pages with
initial in-degree greater than the cutoff degree, but final in-degree less than 1K) for
different cutoff degrees; the upward-slope line denotes the number of \emph{talented winners} (i.e.,
pages with initial in-degree less than the cutoff and final in-degree greater than 1K) as
a function of the cutoff.  For comparison sake, we pick the critical cutoff degree that
equalizes the number of talented winners and experienced losers. Inserting a dashed vertical line
denoting the critical cutoff degree into Fig. (a), we find that 48\% of the winners are talented
winners.
} \label{fig:sep}
\end{figure}

The sequential time-sampled data helps us in better understanding the interplay between
experience  and talent (fitness).   For example, the \emph{initial in-degree of a page}
(i.e., in June 2006) is a measure of its \emph{experience}, and \emph{the accumulated
final in-degree} (i.e., in June 2007) is a measure of \emph{how it fared} based on its
fitness and its experience. We \emph{define a page to be a winner} if its final degree
exceeds a specific desired target, while starting with an initial degree less than the
target value. Figure~\ref{fig:sep} (a) shows the \emph{initial in-degree distribution} of
all pages such that the initial degree was less than 1000  and the accumulated final
degree greater than 1000. The case of different target final degree values is discussed in
Supporting Information. If the growth of the number of hyperlinks acquired by a
page was based purely on PA (i.e., all pages have the same talent/fitness), then only
pages with initial degree greater than a certain threshold would end up with final degree
greater than a thousand. Clearly, the empirical data shows that it is not the case: There
are {\em talented winners} who have very low initial in-degree  and yet end up as winners;
similarly, there are {\em experienced losers} who start with a large in-degree (i.e.,
greater than the cut-off) but yet end up with cumulative in-degree less that 1K.
Figure~\ref{fig:sep} (b) shows the number of talented winners and experienced losers as a
function of the cut-off, and for the sake of fair comparison, we pick a value for the
cutoff such that the number of talented winners equals the number of experienced losers.
Thus, we find that for this sample set, \emph{the web collectively picked $48\%$ talented
winners}, and displaced an equal number of more experienced pages, thus striking a balance
between talent and experience. As analyzed in Supporting Information, the
percentage of talented winners seems to remain relatively constant as the target degree is
varied.

\begin{figure}[htb]
\centering
\includegraphics[width=3.2in]{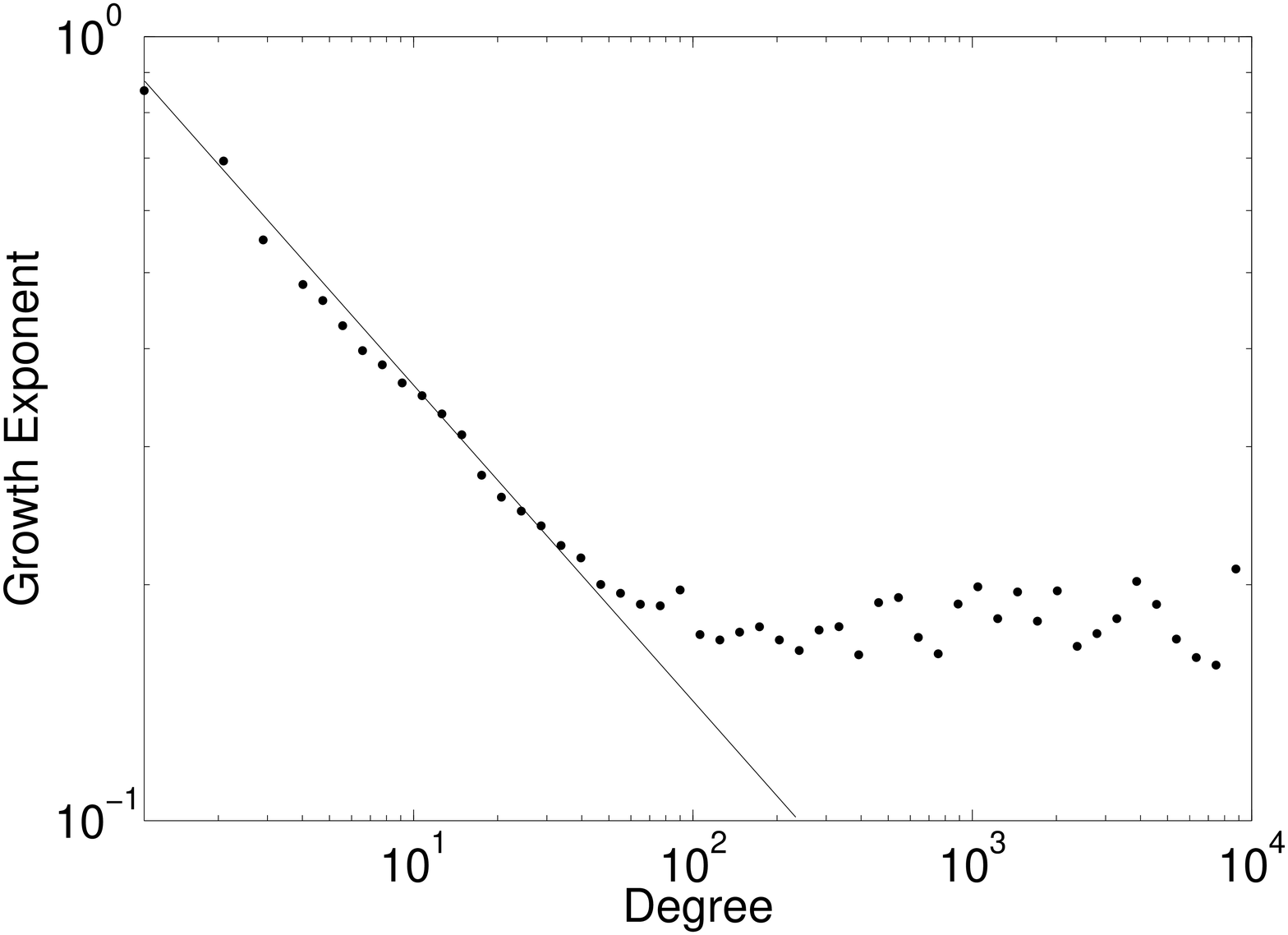}
\caption{\small The average fitness value (as measured by the growth exponent) plotted as a function of the initial in-degree
(or experience) of pages. The set of pages considered consists of those
with measurable fitness (see the {\it Results} section for a
definition of such pages). As the plot shows, pages with low initial in-degree has higher
average fitness, even though the distribution is always exponential; moreover, the average
fitness decreases as a power law form $k^{-0.4}$ until about $k=100$, and then levels off to a
constant value. Thus, the web on the average gives a slight fitness boost to the pages
with low experience, but then treats them statistically the same once they have experience
above a certain value. } \label{fig:avgfitvsdeg}
\end{figure}

\begin{figure}[htb]
\centering
\includegraphics[width=3.2in]{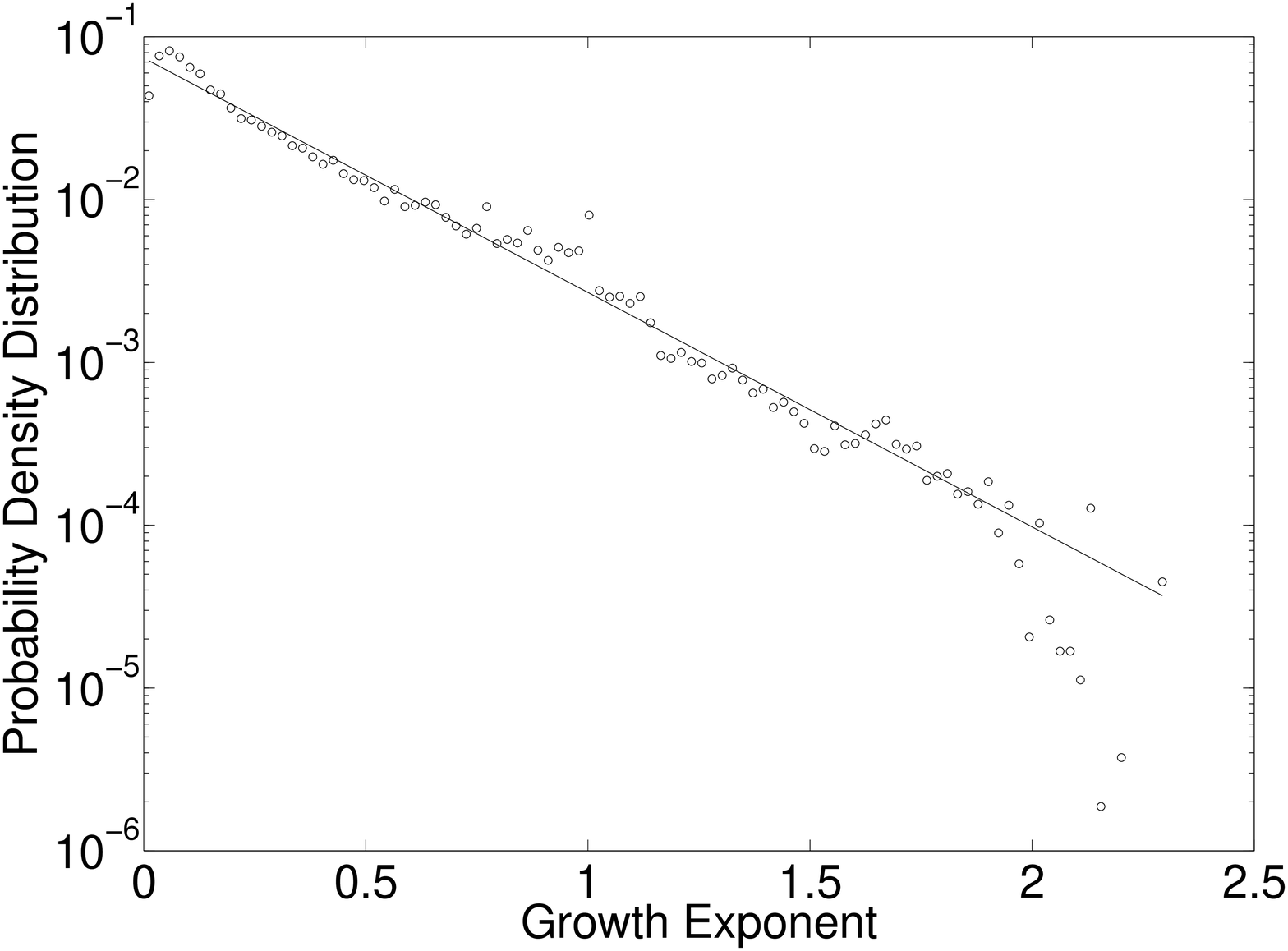}
\includegraphics[width=3.2in]{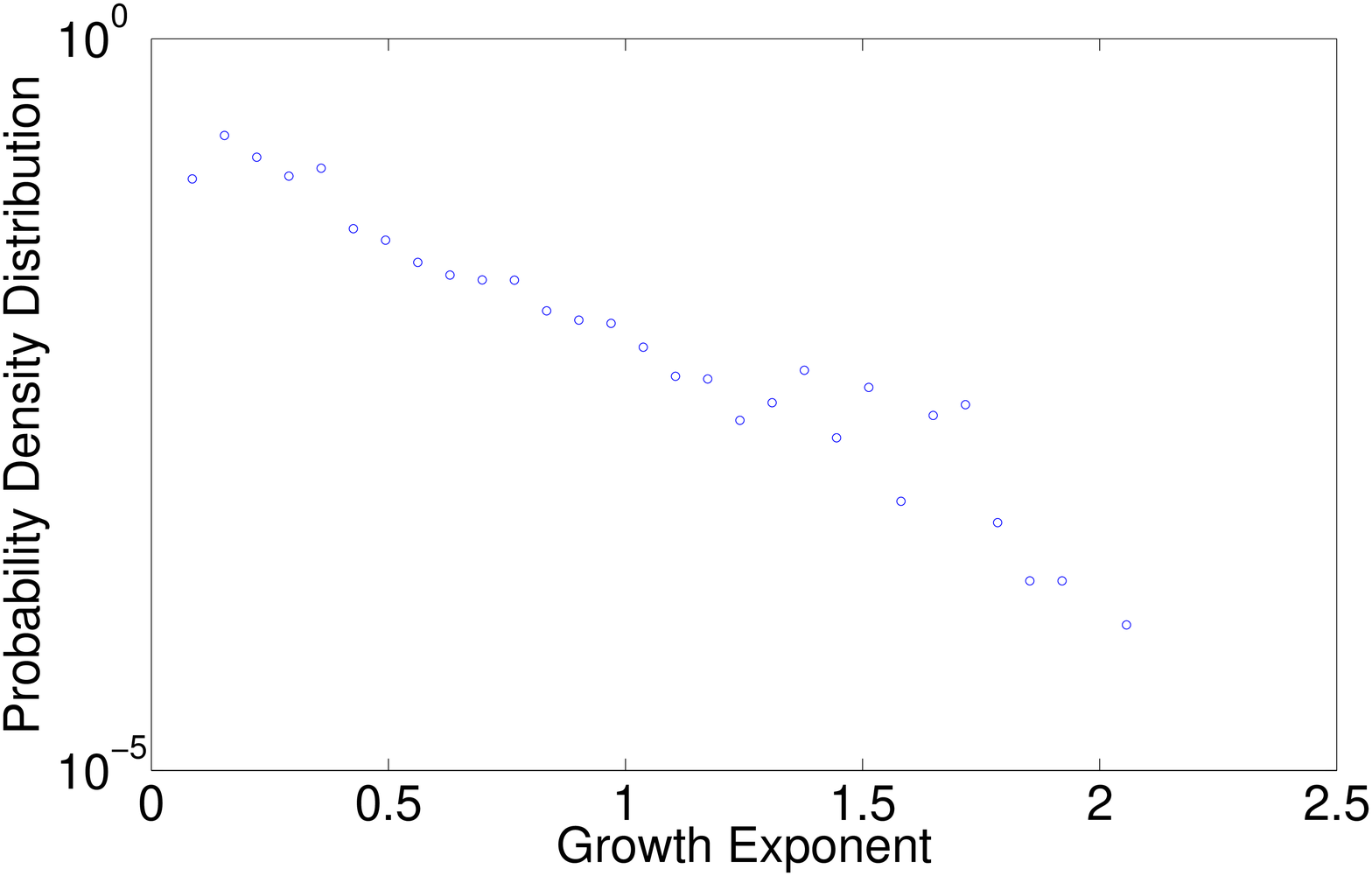}
\caption{\small (a)  Distribution of the growth exponents: the log-linear plot is
well-fitted by a straight line in the range between 0 and 2, which suggests that the
distribution is a truncated exponential. The slope of the fitted line is -1.44.
(b) We find that the growth exponent distribution also exhibits an exponential form when
restricting to sets of nodes with the same initial in-degree in June, 2006; the plot for
the set of nodes with an initial in-degree of $10$ is displayed here. Note that the growth
exponent is an affine function of the underlying fitness parameter (see Eq. \ref{beta});
hence, the fitness distributions are also truncated exponentials. }
\label{fig:web_fit_log}
\end{figure}

What does the fitness distribution look like for pages with similar experience?
Figure~\ref{fig:web_fit_log} (b) shows the fitness distributions of pages with similar
initial in-degree, and hence, similar experience. They all are exponentially distributed,
except that the average fitness is a function of the initial degrees of the nodes.
Figure~\ref{fig:avgfitvsdeg} shows the average fitness as a function of initial in-degree.
It shows that the average fitness is largest for nodes with least experience, and
decreases as a PL until about an in-degree value of 100; it levels off after that.
Thus, the web encourages pages with
low or little experience just a bit more than the mature pages; but for any group, it
judges talent quite conservatively keeping the distribution exponential. 
The concept of fitness has implications on how we rank the importance and attractiveness
of web pages. In the {\it Discussion} section, we propose that one can use the fitness
estimates of the pages to boost their rankings; this way, pages with low overall degree
but that are growing fast will get higher ranking.

\section{Results}

\noindent{\bf Estimating the Fitness of Webpages: Talents Are Exponentially Rare. }
If the fitness with deletions model is indeed applicable to the web, the accumulated degree of each node
should follow Eq.~\ref{k_accum} as discussed in {\it Materials and Methods}.
In particular, from Eq.~\ref{k_accum}, taking the logarithm of both sides of the
accumulated degree of a page, we get:
\begin{equation}
\log k^{*}(i,t) = (\frac{\eta_i}{A}-\frac{c}{1-c})\log t + B = \beta_i\log t + B
\label{eq:k_evol_accum}
\end{equation}
where $B$ is some time-invariant offset. Hence, the slope of the linear fit of the logarithm of the
accumulated degree $k^{*}(i,t)$ and time $t$ gives node $i$'s growth exponent $\beta_i$. Note that the fitness value is related to the
growth exponent of a node by a linear transformation with constant coefficients (see
Eq. \ref{beta}). Thus, the distribution characteristics of fitness can be obtained by
measuring the growth exponent of each node.

The methodology for measuring the distribution of the growth exponents is
described as follows: first, we identify about 10 million webpages that persist through
all 13 months from June 2006 to June 2007. For each of these webpages, the set of
in-neighbors are identified for all months. The accumulated in-degree of a node at any month
is the sum of the in-neighbors up to that particular month. In accordance with
Eq. \ref{eq:k_evol_accum}, after taking the logarithm of the accumulated in-degree and
time (measured in months), the slope of the linear ordinary least-square fit (i.e. the
empirical growth exponent) along with the Pearson correlation coefficient are obtained for
each webpage.  We will refer to this methodology as the growth method; in the Supporting Information,
we present an alternative methodology, the direct kernel method, to estimate the fitness of webpages;
the results from the alternative method is consistent with the results from the growth method.

We found that a large fraction of webpages do not gain any in-connection at all
during the entire 1-year period.  We consider a webpage to have a {\em zero} growth exponent
if its in-degree values increases two times or less during the 13 months.
We found that only 6.5\% of the webpages have {\em nonzero} growth exponents.
We will focus our study on the set of nodes with nonzero growth exponents.
Note that the set of webpages with zero growth exponents essentially introduces a delta function
at the origin in a fitness distribution plot.  It is simple to check that the delta function
does not impact the derivation of results and hence omitted from discussion for simplicity.

An overwhelming fraction of the linear fit produces a correlation coefficient of 0.8 or more,
with an average correlation value of 0.89 (see Supporting Information).
Thus, our empirical measurement is consistent with the model that the evolution of node in-degree as
a function of time follows a power-law as described in Eq. \ref{eq:k_evol_accum}  for majority of  the webpages.

\begin{figure}[htb]
\centering
\includegraphics[width=3.2in]{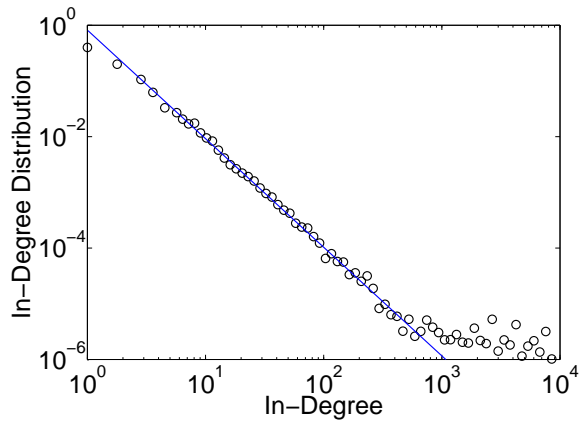}
\caption{\small This figure plots the degree distribution as measured in 2007-06 for webpages that first appear in the month 2006-07.
The fitted power law exponent is $\gamma = 2.0$.}
\label{fig:same_age}
\end{figure}

We plot the distribution of the growth exponents for the set of nodes with correlation
coefficient of 0.8 or more in Fig. \ref{fig:web_fit_log}. The distribution of the growth
exponents has a mean of 0.30 and clearly follows an exponential curve with a truncation
around $2.0$ and a slope of $-1.44$ in the log-linear plot (i.e. a characteristic
parameter of $\lambda=1.44/\log e$). Since node fitness and the growth exponent are
related by a linear transformation involving the constants $A$ and $c$ as $\beta_i = \frac{\eta_i}{A}-\frac{c}{1-c}$, the fitness distribution is also well modeled by the same
form of a truncated exponential.
\\ \\
\noindent{\bf Examples of High-Talent Webpages. } We now conduct checks to see if
the webpages identified to have a large growth exponent indeed contain interesting or important content
that warrants the title of being highly fit or ``talented''. We manually inspected the
several highest-fitness pages in our dataset.  One example is a webpage from the John Muir Trust website that calls on people
to explore nature (http://www.jmt.org/journey). Many in-links to this page is from other sites on nature and outdoor activities.
Another example is the webpage that reports the crime rate of the US from 1960 to 2006
(http://www.disastercenter.com/crime/uscrime.htm). This URL has many in-links from other sites that
discuss different crimes such as murder.
\\ \\
\noindent{\bf Power Law Degree Distribution of the Webpages with the Same Age. }
For scientific citation networks, it is known that the in-degree
distribution of the papers published in the same year follows a power law (see the ISI
dataset in Fig. 1(a) in \cite{redner}). However, no parallel study has been performed for
the web. Using our temporal web dataset, we studied the in-degree distribution of the set
of webpages with the same age. The in-degree distribution is found to follow a power law
with an exponent of $2.0$ for over three decades (see Fig. \ref{fig:same_age}). This
result is consistent with the empirical finding by Adamic and Huberman that the degree
distribution of web hosts with the same age has a large variance \cite{huberman_tc}.
Furthermore, the power law nature of the in-degree distribution is consistent with our
theoretical prediction from Eq. \ref{eq:degdist_sameage} given the fitness distribution
is found to be a truncated exponential (see {\it Materials and Methods}).

In contrast, a network dynamic model that does {\em not} account for fitness has
a small variance for the nodes with the same age, which leads to the effect that the ``rich'' node must be the old node.
In fact, this is the basis of the issue raised by Adamic and Huberman \cite{huberman_tc}.
Thus, {\em the fitness-based model naturally generates the power law degree distribution for the set of nodes
with the same age}, which is not explained by other existing models that do not account for fitness
such as \cite{barabasi,sarshar,chung,cooper,moore}.
\\ \\
\noindent{\bf Ad Hoc Characteristics of the Web and the Resilience of the Power Law Exponent.}
We now discuss the webpage removal process as
observed in our dataset. In our analytical model, a node is removed uniformly randomly
(i.e. independent of node degree). We found empirical evidence to support the uniform
random removal assumption: we observed that the degree distribution of the set of removed
nodes that disappear in a given month is similar to the degree distribution of all nodes
(see Supporting Information).

Recall that the turnover rate is defined as the average number of nodes removed per node
added. From our dataset, the turnover rate is measured to be $c=0.91$ (i.e. for every new
webpage inserted, 0.91 webpage is removed per unit time).  However, this figure is an
overestimate of the true turnover rate on the web, since we are examining a fixed set of
web hosts. Therefore, we also need to account for the growth in the number of web hosts.
Nevertheless, even after accounting for the source of growth from the insertion of web
hosts, web still has a minimum turnover rate of 77\% (see Supporting Information).

\begin{figure}[htb]
\centering
\includegraphics[width=1.5in]{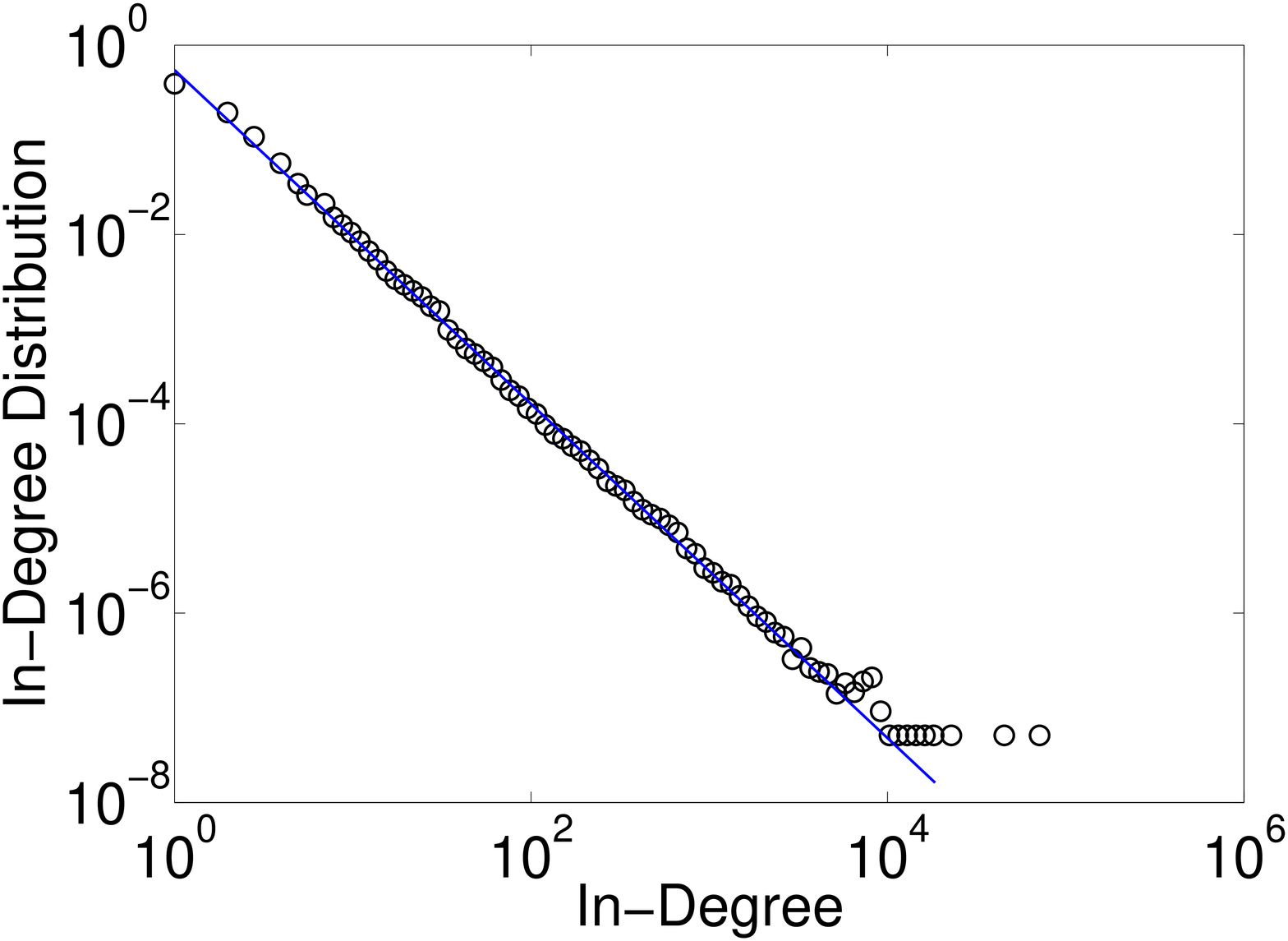}
\includegraphics[width=1.5in]{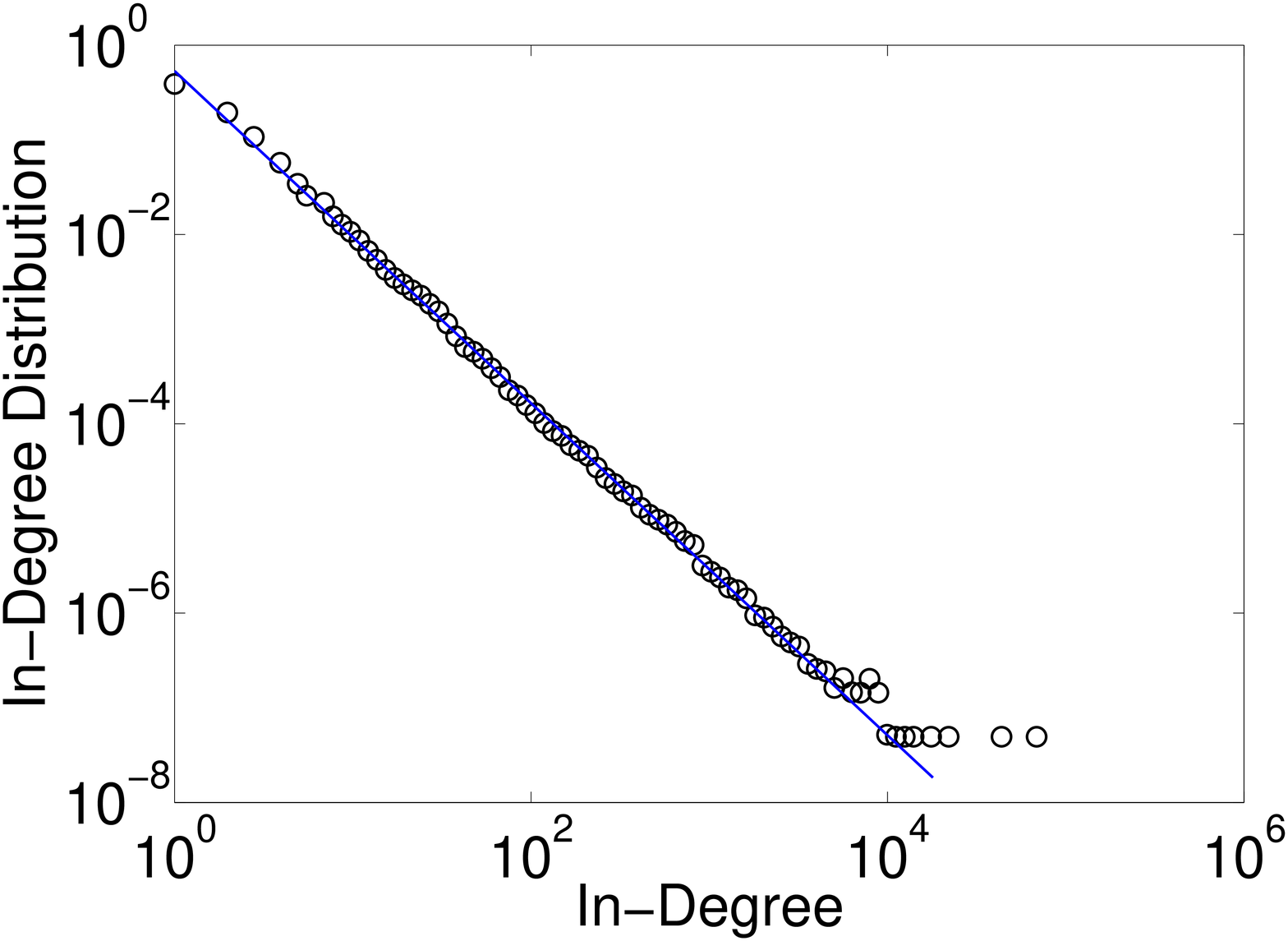}
\includegraphics[width=1.5in]{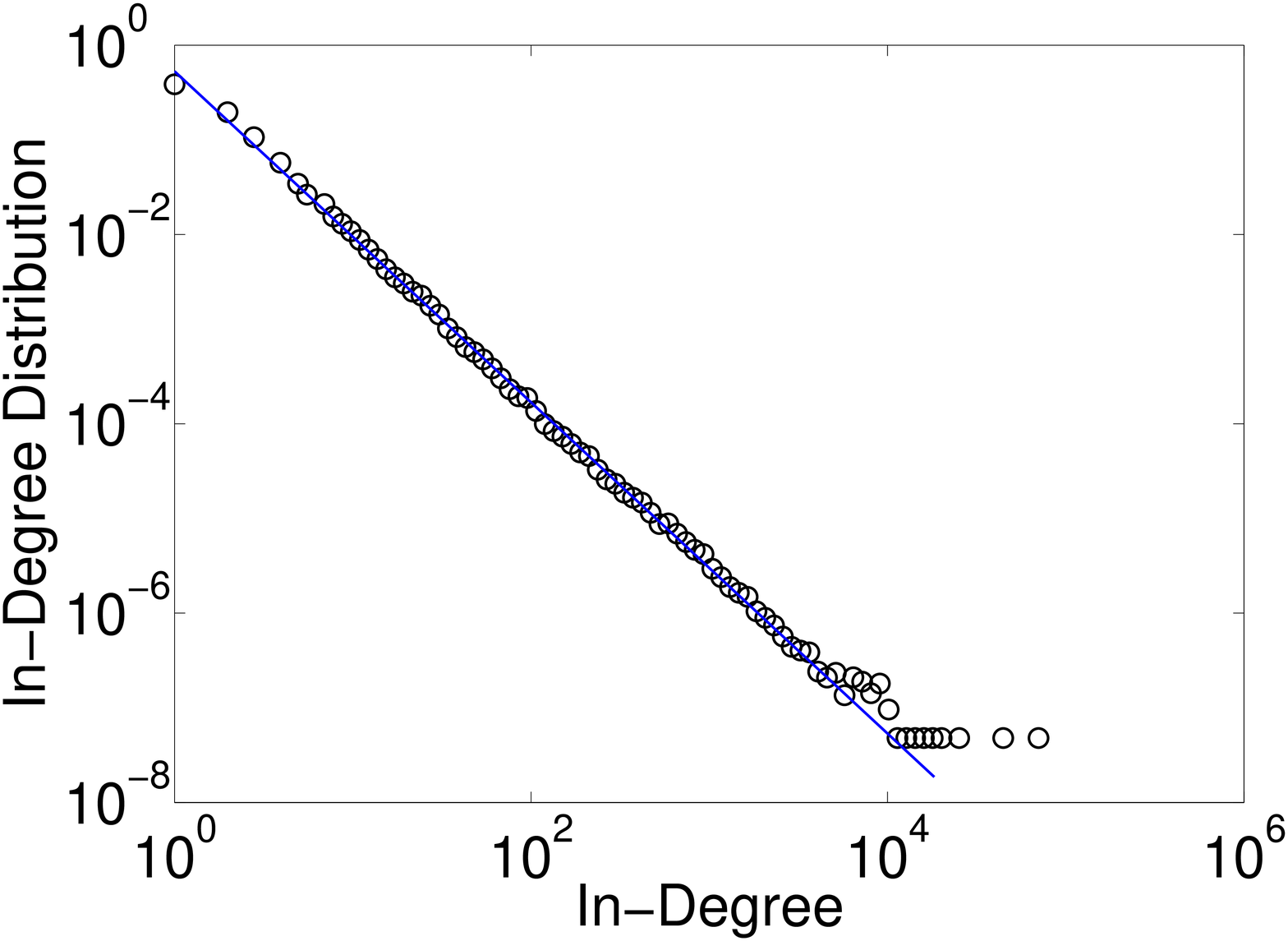}
\includegraphics[width=1.5in]{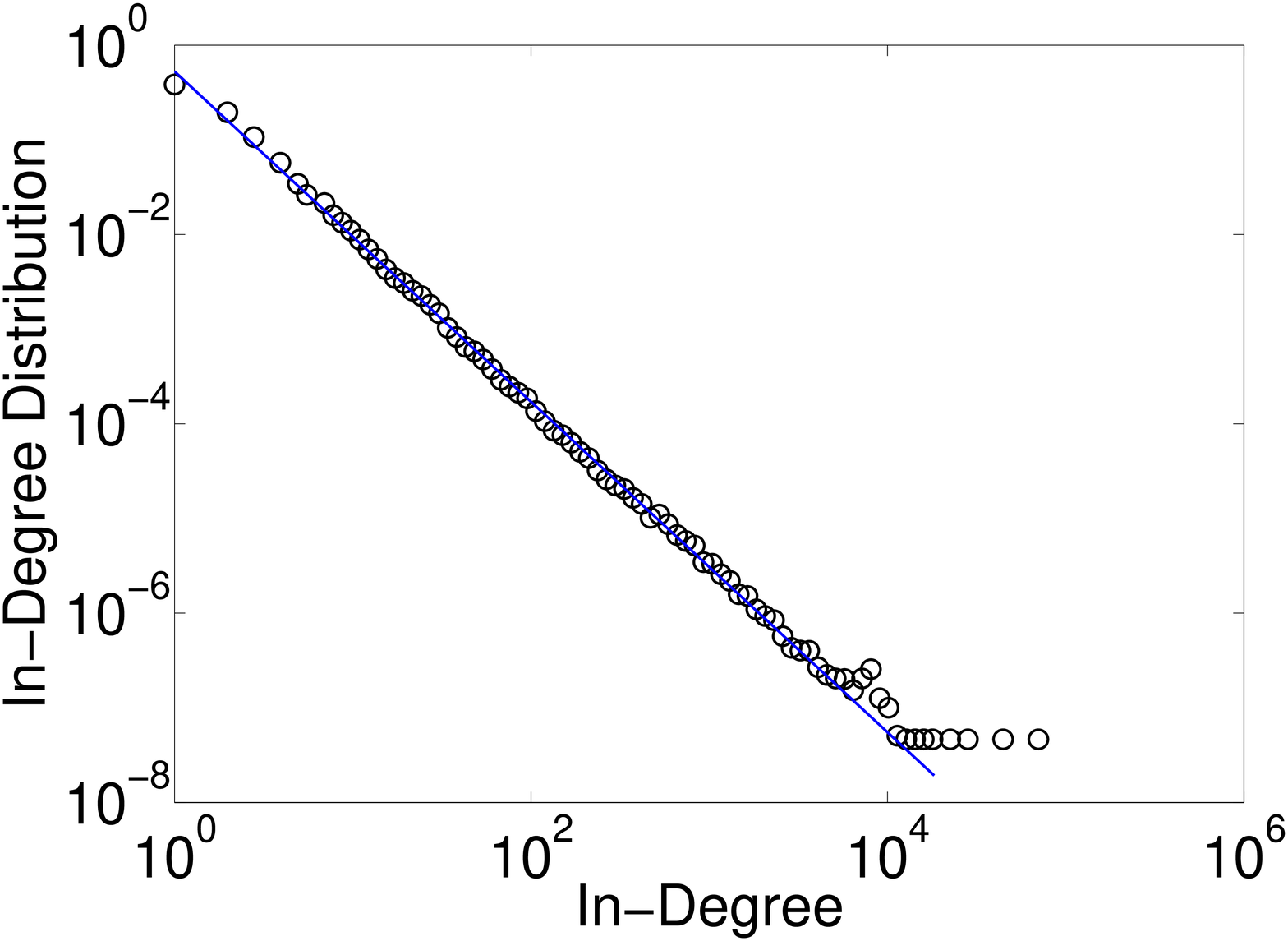}
\caption{\small In-degree distributions for different months: 06-2006, 10-2006, 02-2007 and 06-2007.
All distributions decay as a power law with the exponent $\gamma_{in}=1.8$
over four decades.  }
\label{fig:gamma_month}
\end{figure}

Despite the high rate of node turnovers, the power law degree distribution  is found to be
very stable (see Fig. \ref{fig:gamma_month}). This finding is consistent with our ad hoc
fitness model prediction that the power law exponent $\gamma$ of the degree distribution
$P(k)\propto k^{-\gamma}$ stays constant for varying rates of node deletion for a
truncated exponential fitness distribution (see Eq. \ref{gammais2} in {\it Materials and Methods}).  The resilience of
the power law exponent is in stark contrast to the result obtained for the
PA-with-deletion model (without any fitness variance), where the power law exponent is
found to diverge rapidly as $\gamma=1+2/(1-c)$ \cite{sarshar,moore}. Thus, {\em the
natural variation of node fitness provides a self-stabilization force for the power law
exponent of the degree distribution under high rate of node turnovers}. \\ \\
\noindent{\bf Talented Winners versus Experienced Losers. }
In the {\it Introduction}, we proposed the idea of talented winners and experienced
losers and how they are identified in our empirical web dataset for a given target degree
$k_{tg}$. For the particular case of $k_{tg}=1000$, we find that 48\% of the winners are
talented winners (see Fig. \ref{fig:sep}), who successfully displaced the experienced
losers (i.e. the nodes with higher initial in-degrees but fail to become a winner). This
observation is seemingly paradoxical: how can talents emerge to win close to half of the
times when talents are exponentially rare? We seek to understand the interplay between
experience and talent through analytical modeling.

Consider a node with the initial degree $k < k_{tg}$ in month $1$ (i.e. June, 2006, the
start of our observation period). In order for the node to achieve $k_{tg}$ in month $13$
(i.e. June, 2007, the end of our observation period), the growth exponent of the node must
exceed the critical value: $\beta_c(k) = \frac{\log\frac{k_{tg}}{k}}{\log 13}$. The
fraction of nodes that are winners is simply given as:
\begin{equation}
W(k_{tg})=\int_1^{k_{tg}} C(\beta_c(k))P(k) dk \label{eq:frac_win}
\end{equation}
where $C(\beta)$ is the complementary cumulative distribution function (CCDF) of the
growth exponent distribution and $P(k)$ is the initial degree distribution in month $1$.
Thus, one can find the fraction of winners for a given $k_{tg}$ by performing numerical
integration of Eq. \ref{eq:frac_win}.

We now introduce the cutoff $k_{cut}$: the set of winners with an initial degree $k < k_{cut}$ are denoted
as the {\em talented winners}, since they start with a low initial degree but nevertheless reach the
target degree $k_{tg}$ in month $13$; the set of losers with an initial degree $k > k_{cut}$ are denoted
as the {\em experienced losers}, since they start with a high initial degree but still fail to reach the
target degree $k_{tg}$ in month $13$.  We can solve for the critical cutoff $k_{cut}^*$ such that the number
of talented winners $TW(k_{cut}^*)$ is equal to the number of experienced losers $EL(k_{cut}^*)$
(i.e. the talented winners displace the experienced losers):
\begin{equation}
TW(k_{cut}^*) = \int_1^{k_{cut}^*} C(\beta_c(k))P(k) dk =  \int_{k_{cut}^*}^{k_{tg}}
(1-C(\beta_c(k)))P(k) dk = EL(k_{cut}^*) \label{eq:kcut}
\end{equation}
The above equation can be solved numerically to obtain $k_{cut}^*$.  Now, the fraction of
talented winners or experienced losers is simply given by: $r_{tw}(k_{tg}) = TW(k_{cut}^*)/W(k_{tg})$.

From our empirical web data, the growth exponent of the nodes is distributed according to
a truncated exponential function $C(\beta)$ with the parameter $\lambda=1.44/\log e$ and
the truncation $\beta_{max}=2.0$. The initial degree distribution $P(k)$ is a power law
with exponent $\gamma=1.8$. Substituting the empirically obtained $C(\eta)$ and $P(k)$
functions into Eq. \ref{eq:frac_win} and \ref{eq:kcut}, we use numerical integration to
find the fraction of talented winners $r_{tw}(k_{tg})$ for the target degree $k_{tg}=1000$ and
obtain the theoretical prediction of 48.8\%, which matches well with the empirical measurements obtained as described in Fig. \ref{fig:sep}
(See Supporting Information for theoretical and measurement results for different target degrees).
For a given system with known talent and initial degree distribution, one can now estimate the
fraction of talented winners using our analytical model.

\section{Discussion}

The competition between experience and talent arises in all aspects of society on a
frequent basis: from choosing an applicant to fill a highly coveted job to deciding which
political candidate to vote for.  Although the study of the interplay between experience
vs. talent has long interested scientists and investigators, and much on the topic has
been theorized about \cite{huberman,bianconi,borgs}, large-scale empirical study on this
topic from a quantitative perspective has been lacking.

In this paper, taking advantage of the large, open and dynamic nature of the World Wide
Web, we find an intricate interplay between talent and experience. Talents are empirically
found to be exponentially rare.  However, through empirical measurements and theoretical
modeling, we show that the exponentially distributed talent accounts for the following
observed phenomena: the heavy-tailed power law in-degree distribution of the web pages
born at the same time, the preservation of the low power law exponent even in the face of
high rates of node turnovers, and most intriguing of all, talented winners  emerge and
displace the experienced losers in just slightly less than half of all winning cases!

Beyond the interesting findings, we discuss several issues associated with this work.
While our data is statistically consistent with the model assumption of a constant fitness
for each page, our observation period is over a relatively short period of one year. For
longer periods, one would expect the fitness of a page to change. For example,
occasionally, a page that has been lying dormant for a while might find its content become
topical and, hence, its fitness suddenly increases, allowing it to start accumulating
links and becoming popular. Such pages can be referred to as sleeping
beauties~\cite{simkin:theory}. Developing a model
that accounts for time-varying fitness can be a subject for future work. In addition, the
sample size on the order of tens of millions of nodes used in this study is arguably
large, especially in comparison to studies from the social sciences. However, the size of
the web is currently on the order of billions of pages. Nevertheless, the source of our
data, the Stanford WebBase project, to the best of our knowledge is the largest publicly
accessible web archive available for research studies. Finally, the statistics on node in-degrees as reported in
this work is measured from the crawled web graph; potential in-links from webpages not included
in the crawl are not accounted for.  Future work focusing on examining larger web samples
can mitigate these limitations.

On the World Wide Web, the problem of search engine bias or the ``entrenchment effect''
(i.e. the ``rich-get-richer'' mechanism) has received considerable attention from a broad
audience from the popular press to researchers \cite{cho_impact}. However, researchers
have shown evidence that the ``rich-get-richer'' mechanism might be less dominant than
previously thought \cite{pennock,fortunato}; nevertheless, search engine bias and the
``entrenchment effect'' remains a concern. The findings in this paper present an
alternative perspective on this problem and show that talents, while being exponentially
rare, are frequently afforded the opportunity to overtake more ``entrenched'' web pages
and emerge as the winner.

Currently, for any given query, pages are ranked based on a
number of metrics, including the relevancy score of the query keywords in a document, and
the document's pagerank, which is computed based on the in-degree (or experience) of the
page and the hyperlink structure of the web at the time of the crawl. In order to avoid
the entitlement bias potentially introduced due to pagerank, a number of researchers have
advocated that one should also boost low pagerank pages, for example, by randomly
introducing them among the top pages \cite{cho_quality,pandey}. The fitness of a page
could be added as another metric that could influence the ranking.  The determination of
the exact functional form of how the fitness, $\eta_i$, of a page would influence its rank
would require considerable experimentation and editorial evaluations, but a promising
start would be to multiply the currently computed ranks by $\eta_i^\alpha$, where the
exponent $\alpha$ is tuned based on quality assessment and testing. This would allow users
to find pages that do not have high page rank yet, but are catching up fast. We expect
such fitness-based ranking algorithms to have widespread applications beyond the web in
other domains
that employ ranking algorithms.  
We will note in passing that as with any other ranking algorithm based on link structure,
the proposed ranking scheme must be used in conjunction with link farm detection
algorithms to minimize the effect of link spamming that might try to influence the
estimation of the fitness factors.

Besides the web, the methodologies developed in this work is applicable for studying other
complex networks and systems such as the citation network of scientific papers and the actor collaboration social network,
where the interplay between ``experience'' and ``talent'' is also interesting.
The fitness distribution is arguably an important parameter for dynamically evolving networks.
The empirical study and theoretical models presented in this paper pave the road for studying the fitness
characteristics of other systems, which will allow us to better understand, characterize and model a broad range of
networks and systems.

\section{Materials and Methods}

\noindent{\bf Dataset. }
Our dataset of the World Wide Web was obtained from the Stanford WebBase project. WebBase
archives monthly web crawls from 2006 to 2007.  We downloaded a total of 13 crawls for a
one year period from June 2006 to June 2007. These crawls track the evolution of  17.5
thousand web hosts with each crawl containing in excess of 22 million
webpages\footnote{The WebBase crawler would extract a maximum of 10 thousand pages per
host. However, the 10 thousand pages per host limit is not a problem, since none of the
page count of the tracked hosts reaches this limit.}. The set of hosts consists of a
diverse sample of the web: it contains 5.4 thousand .com hosts, 4.7 thousand .org hosts
and 2.6 thousand .edu hosts.  This set also includes many foreign hosts, such as hosts
from China, India and Europe.
\\ \\
\noindent{\bf A Fitness-based Model for Ad Hoc Networks.}
The existing ``preferential attachment with fitness model''  is specified as follows
\cite{bianconi}: at each time step, a new node $i$ with fitness $\eta_i\geq 0$ joins the
network, where $\eta_i$ is chosen randomly from a fixed fitness distribution $\rho(\eta)$;
node $i$ joins the network and makes $m$ links to $m$ nodes. A link is directed to node
$l$ with probability:
\begin{equation}\label{eqn:kernel}
\Pi_l=\frac{\eta_l k_l}{\sum_j \eta_j k_j}
\end{equation}
where $k_l$ is the in-degree of the node $l$.    We extend the fitness model to account for node
deletion. The new model, which may be called  ``fitness with deletion model'', has the
following extra dynamic added to the original fitness model: at each time step, with
probability $c$, a randomly selected node is deleted, along with all of its edges. We
present the analysis of the model using the continuous mean-field rate equation approach
as introduced in \cite{dorogov_scaling}. Other approaches would include the generating
function method as discussed in \cite{moore} and the rigorous mathematical analytical
method presented in \cite{aiello:focs}. However, we prefer the mean-field approach for its
simplicity. In addition, the analytical results are verified by simulations.
On another note, since the web is a directed graph, we note that the model can be easily generalized into a
dynamic directed network model (details are discussed in the Supporting Information).

In the fitness with deletion model, we show that the evolution of $k_i$ follows a power-law (see Supporting Information):
\begin{equation}
k_{\eta_i}(i,t)=m \left( \frac{t}{i} \right)^{\beta(\eta_i)}, \label{power}
\end{equation}
where the {\em growth exponent} $\beta$ is a function of the fitness $\eta_{i}$:
\begin{equation}
\beta(\eta)=\frac{\eta}{A} - \frac{c}{1-c} \label{beta}\ .
\end{equation}
The parameter $A$ is given by:
\begin{equation}
1= \int_0^{\eta_{max}} d\eta \rho(\eta) \frac{1}{\frac{A}{\eta}\frac{1+c}{1-c}-1} \label{integral}
\end{equation}
where $\eta_{max}$ is the maximum fitness in the system.

We now examine the case where the fitness distribution is a truncated exponential, which is shown to empirically
characterize the fitness distribution of webpages. When $\rho(\eta)$ is distributed
exponentially in the interval $[0,\eta_{max}]$, we have: $\rho(\eta) = \lambda
e^{-\lambda\eta}/(1-e^{-\lambda\eta_{max}})$. The constant $A$ can be determined from Eq.
\ref{integral}. For $\eta_{max}$ large compared with 1/$\lambda$, we have
$A=(\eta_{max}+\epsilon_1)\frac{1-c}{1+c}$, where $\epsilon_1$ is negligibly small. Thus,
according to Eq. \ref{beta}, the growth exponent is given by
\begin{equation}
\beta(\eta) = \frac{\eta(1+c)}{(\eta_{max}+\epsilon_1)(1-c)}-\frac{c}{1-c}
\end{equation}
For maximum fitness $\eta=\eta_{max}$, we have $\beta_{max} = \frac{1-\epsilon_2}{1-c}$,
where $\epsilon_2$ is negligibly small.  Since the power law exponent is dominated by the highest $\beta$,
we invoke the {\em scaling relation} \cite{sarshar} $\gamma(\eta) = 1+\frac{1}{(1-c)\beta(\eta)}$; we obtain:
\begin{equation}
\gamma = 2+\epsilon_3. \label{gammais2}
\end{equation}
where $\epsilon_3$ is negligibly small.  Thus, the power law exponent stays at 2
regardless of the deletion rate (see Supporting Information for the detailed derivations
and justifications on assumptions made). This is a rather surprising result. As was shown
in \cite{sarshar,chung,cooper,moore}, for plain preferential attachment dynamics (where
all nodes have the same fitness), the power-law exponent depends on $c$ as
$\gamma=1+2/(1-c)$, and diverges as $c$ goes to 1. The introduction of fitness with a
truncated exponential distribution \emph{stabilizes} the power-law exponent, in the sense
that the exponent remains close to 2.0 and does not diverge, regardless of the value of
$c$. To verify the result that the power-law exponent does not depend on the turnover
rate, we performed large-scale simulations and confirmed that the power-law exponent stays
constant at 2.0 even under high rates of node turnovers (see Supporting Information).
\\ \\
\noindent{\bf Degree Distribution of Nodes with the Same Age. }
Given the fitness distribution $\rho(\eta)$ and the degree, $k$, that grows exponentially
with fitness for a fixed time interval $t$, we have $k(\eta) \propto t^{\eta/C}$, where
$C$ is some constant.  The degree distribution of nodes with the same age is given as:
$P(k) = \rho(\eta)\frac{d\eta}{dk} \propto \rho(\eta)\frac{C}{k\ln t}$.
For the case that the fitness distribution is a truncated exponential $\rho(\eta) \lambda e^{-\lambda\eta}/(1-e^{-\lambda\eta_{max}})$, the degree distribution follows a power law:
\begin{equation}
P(k) \propto \frac{\lambda e^{-\lambda \eta}}{1-e^{-\lambda \eta_{max}}}\frac{C}{k\ln t} \propto k^{-\gamma} \label{eq:degdist_sameage}
\end{equation}
where the power law exponent is $\gamma = 1+\frac{\lambda}{\ln t}$. Effectively, the
light-tailed distribution in fitness is {\em amplified} into the heavy-tailed degree
distribution for nodes born at the same time through the PA mechanism. The phenomenon of
heavy-tailed degree distribution of nodes with the same age has also been observed and
analyzed in other contexts \cite{redner:1998,simkin:aces,simkin:theory}.
\\ \\
\noindent{\bf The Evolution of the Accumulated Node Degree. }
In our model, a node would gain neighbors as well as lose neighbors when the neighboring
nodes are deleted.   As a result, when we track the evolution of a node's degree over
time, the time series shows a number of upward and downward jumps, making it difficult to estimate the growth
exponent $\beta(\eta)$ from Eq.~\ref{power} accurately.
In order to reduce noise in the data, we can instead track the evolution of a
node's {\em accumulated} degree over time. We  define the set of accumulated neighbors of
a node to include previous neighbors that have been deleted in addition to the current set
of neighbors. Thus, the accumulated node degree is the size of the set of accumulated
neighbors. It is simple to derive that the evolution of the accumulated degree of node $i$
is (see Supporting Information):
\begin{equation}
k_{\eta_i}^{*}(i,t) = m\frac{\eta_i(1-c)}{\eta_i(1-c)-cA}\left(\frac{t}{i}\right)^{\beta^{*}(\eta_i)}\label{k_accum}
\end{equation}
where the growth exponent is found to be $\beta^{*}(\eta)=\frac{\eta}{A} - \frac{c}{1-c}$.
Note that the growth exponent for the evolution of the accumulated node degree $\beta^{*}(\eta)$ is
identical to the growth exponent of node degree as given in Eq. \ref{beta} (i.e. $\beta^{*}(\eta) = \beta(\eta)$).

\section{Acknowledgement}

The authors would like to thank Gary Wesley of the Stanford WebBase project for providing patient help and instructions in downloading the web crawl data.


\mbox{ }
\newpage
\mbox{ }

\end{document}